# SciBlog : A Tool for Scientific Collaboration

L.T. Handoko[1]

Group for Theoretical and Computational Physics[2], Research Center for Physics LIPI
Kompleks Puspiptek Seprong, Tangerang 15310, Indonesia

*Abstract*

I describe a newly developed online scientific web-log (SciBlog). The online facility consists of several moduls needed in a common and conventional research activity. I show that this enables scientists around the world to perform an online collaboration over the net.

## 1. Introduction

Research activities are, by nature, done in a global environment since long time ago, much longer than globalization in economics and markets. Scintific progress has been realized as a result of universality, equal-treatment and non-discrimination of scientific works. This background boosts collaborations among multi people, institutions and even multi nations. This nature actually accelerates scientific progress faster day by day. Because scientific contributions are explored from infinite number of peoples without interfered by their backgrounds. As a result, human-being enjoy advanced technologies surrounding their lifes at time being. On the other hand, a rapid scientific progress in a relatively short period makes the competition between the scientist is getting tighter, and in some cases is getting hard to follow. These scientific requirements, in both quantity and quality, again encourage scientists around the world to work together and collaborate each other.

Unfortunately, collaboration is in most cases very costy. It requires enough and non-trivial financial supports. It is also very time consuming to make travel and even communicate each other. More than that, life discussion and communication among the members is unrecordable. This point is very crucial especially for a reseach collaboration which is application-oriented, and the final result is, for instance, a registered patent. Also, a complete record is important as a tool to evaluate the contributions of each member in a collaboration group, which can be used to determine a fair capital share later on if the undergoing research project is successful.

According to these backgrounds, it is then very important to start thinking how we can manage the knowledge generated in a research collaboration. This is the main motivation to develop the full online scientific web-log, later I call it in short as SciBlog [1]. First I will describe briefly the system and topology, followed by explaining its present status and future prospects before ending with the summary.

---

1  handoko@lipi.fisika.net, handoko@fisika.lipi.go.id
2  http://www.fisika.lipi.go.id

## 2. System and topology

Now, we already understood how important the knowledge management system. The problem is how we can implement it under special circumstances typical in developing countries which are in most cases lack of financial and infrastructure resources. Of course the system is not intended to be specific only for communities in the developing countries, but once the system works for them then it would be automatically applicable for the developed countries as well. Because the developing countries have much more limitations than its counterparts.

In order to make the system works well and acceptable for the scientific communities, especially in the developing countries, we have developed the system with the following basic principles :
– The system should be able to be maintained and used by all parties (the host, the owners of a group and the end-users) inexpensively, or if possible at no-cost.
– The system should accommodate all aspects in a scientific collaboration normally done in a conventional manner.
– The system should be easy to use (user-fiendly) for all levels of potential users.

We have been working hard to fullfill all of either those requirements or limitations. First of all we put an assumption that the system will be run and hosted by a non-profit body like Indonesion Institute of Sciences (LIPI). We have finally decided that the system will be maintained, after a development period, by the focal point of CODATA (Committee on Data for Scientific and Technology) in Indonesia [2] which is hosted by LIPI. CODATA Indonesia actually has been maintaining all IT-based public services owned by LIPI since few years ago [3].

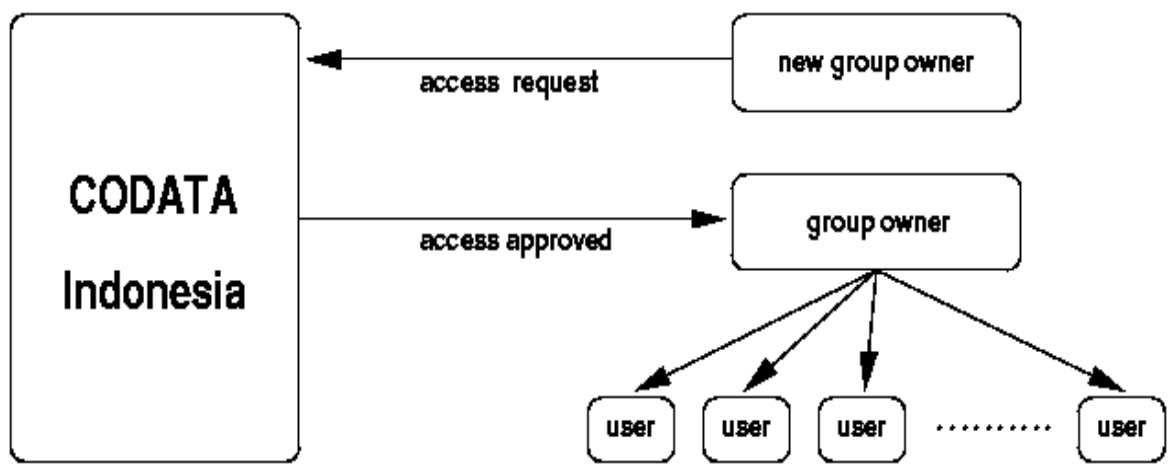

Figure 1. Topology of the users (the host, the group owners and the end-users) in SciBlog.

The second point is relatively easy to realize since by nature there is only limited bureaucracy in an research collaboration, and recently everything is done on digital based works. We have just collected all aspects and integrated them into a compact application accessible by clicking its links on the web. Basically, we have merged the functionalities of mailing-list and web-log which are already familiar for most of internet users. Of course, in order to improve the functionality we have introduced in

addition to both well-known facilities :
- Customizable web-page for the group and its unique domain name. This can be used to advertise its activities and provide related information for public.
- Closed discussion forum that is limited only for the member of the group. This keeps live discussion among the members to be confidential.
- Digital data shared to the members. The data could be everything : figure, sound, image, video, text and so on.
- Database for the publications and related results generated during the collaboration. This database will also be accessible for public through the web-page mentioned above.
- Balance sheet shared to the members. This is crucial to enable a transparant management for a collaboration project.
- Agenda shared to the members contains all schedules related to the collaboration.
- Favourite links shared to the members.
- Internal message system as communication tool among the members and from the external visitors as well. This message system is equipped with an alert system to inform the (internal) senders if theirs messages has been read by the recipients.

Beside those moduls, each groups has also a user management system which enable the group owners to maintain their members and set the priviledges for each of them. The topology and flow-chart of all users incorporated in the system is shown in Fig. 1.

The last point is probably the most important one, otherwise the system is meaningless. Considering the limitations especially in the developing countries, we have choosed the widely used *http* protocol. This means all users should be able to access all modules through the web using any browsers. We have used the Linux [4] as the operating system and the most reliable web-server, Apache [5]. Limiting the size of each web-page is also urgent to facilitate the potential users with low bandwith, *i.e*. The users with dial-up connection of 33 Kbps. We have put an upper bound for the web-page dynamically generated in each session to be less than 25 Kb. This tight requirement can be realized by using Perl code [6] in all modules, and we have developed our own database rather than used generally available databases such as SQL etc. All of these make the system is cheap to deploy, very light and speedy even installed in a server with the specification of personal PC.

## 3. Present status and future prospects

The system is alrealy in full run since September 1$^{st}$ 2004, installed in a server with dual Xeon processors hosted at the main NOC of PT Indosat [7] which owns the main internet gateway from and to Indonesia.

At the time of writing this processding, the system has been used by 9 research collaborations around Indonesia, mainly Jakarta. It shows that the system still needs more publicity and we have to pay some efforts to spread it to the potential users in Indonesia. As the number of user is growing we do hope to obtain more valuable feedback needed to either fix the existing tools or improve new tools.

In the future, the system has a potential to be used by global scientific coomunities around the world. When the time is coming, we plan to develop an English version for the international users.

## 4. Summary

The newly launched SciBlog has been introduced and explained briefly. All potential and present problems has been clarified. I do hope the readers can make a try as the first wave of potential users.

**Acknowledgement**
This work is supported by the fund of DIP Kompetitif LIPI – Teknologi Informasi (contract no. 378378). I also thank the Organizer of Workshop on Knowledge Management 2004 at Bedugul Bali, Indonesia for their warm hospitality during the workshop.